\documentclass{article}
\usepackage{amssymb}
\usepackage{amsfonts}
\usepackage{amsmath}
\usepackage[english]{babel}
\usepackage{bbold}
\usepackage{cite}

\begin{document}

\title{Generalized solutions of the Dirac equation, \\
W bosons, and beta decay}
\author{Andrzej Okni\'{n}ski\thanks{
Email: fizao@tu.kielce.pl} \\
%EndAName
Chair of Mathematics and Physics, Politechnika \'{S}wi\c{e}tokrzyska, \\
Al. 1000-lecia PP 7, 25-314 Kielce, Poland}
\maketitle

\begin{abstract}
We study the $7\times 7$\ Hagen-Hurley equations describing spin $1$
particles. We split these equations, in the interacting case, into two Dirac
equations with non-standard solutions. It is argued that these solutions
describe decay of a virtual $W$ boson in beta decay.
\end{abstract}

\section{Introduction}

\label{introduction}

Recently, we have shown that in the free case covariant solutions of the $%
s=0 $ and $s=1$ Duffin-Kemmer-Petiau (DKP) equations are generalized
solutions of the Dirac equation \cite{Okninski2015b}. These wavefunctions
are non-standard since they involve higher-order spinors. We have
demonstrated recently that in the $s=0$ case the generalized solutions
describe decay of a pion \cite{Okninski2015a}. The aim of this work is to
interpret spin $1$ solutions, possibly in the context of weakly decaying
particles.

There are several relativistic equations describing spin $1$ particles, see 
\cite{Beckers1995a,Beckers1995b} for the reviews. The most common approach
to study properties of spin $1$ bosons is based on the $10\times 10$ DKP
equations (the DKP particles are bosons \cite{Bennett2016}). Several classes
of potentials were used in DKP equations to investigate interactions of spin 
$1$ particles \cite%
{Kozack1989,Mishra1991,Nedjadi1994,Cardoso2010,Molaee2012,Hassanabadi2012,Hassanabadi2013a,Hassanabadi2013b, Castro2014,Hassanabadi2014}%
. However, we shall apply the $7\times 7$\ Hagen-Hurley equations \cite%
{HagenHurley1970,Hurley1971,Hurley1974} in spinor form \cite%
{Okninski2015b,Lopuszanski1978,Dirac1936}. Our motivation stems from the
observation that these equations violate parity and thus should describe
weakly interacting particles.

In the next Section we transform the Hagen-Hurley equations, in the
interacting case, into two Dirac equations with non-standard solutions
involving higher-order spinors, extending our earlier results described in 
\cite{Okninski2015b}. These generalized solutions bear some analogy to
generalized solutions of the Dirac equation argued to describe a lepton and
three quarks \cite{Marsch2015}. In Section \ref{betaW} we describe
transition from non-standard solutions of two Dirac equations to the Dirac
equation for a lepton and the Weyl equation for a neutrino. In the last
Section we show that the transition is consistent with decay of a virtual $W$
boson in beta decay. In what follows we are using definitions and
conventions of Ref. \cite{Okninski2012}.

\section{Generalized solutions of the Dirac equation in the interacting case}

\label{generalized}

We have shown recently that, in the non-interacting case, solutions of the $%
s=0$ and $s=1$ DKP equations are generalized solutions of the Dirac equation 
\cite{Okninski2015b}. In our derivation we have splitted the $10\times 10$
DKP equations for $s=1$ into two $7\times 7$\ Hagen-Hurley equations \cite%
{HagenHurley1970,Hurley1971,Hurley1974}. Let us note here that in the case
of interaction with external fields such splitting is not possible since the
identities (27) of Ref. \cite{Okninski2004}, enabling the splitting, are not
valid in the interacting case. Therefore, we shall base our theory on the $%
7\times 7$\ formulation, see Eqs. (18), (19) in \cite{Okninski2015b} and
Subsection 6 ii) in \cite{Lopuszanski1978}. These equations violate parity $%
P $, where $P:x^{0}\rightarrow x^{0},\ x^{i}\rightarrow -x^{i}\ \left(
i=1,2,3\right) $, and thus one should expect a link with weak interactions.

We write one of these $7\times 7$\ equations (Eq. (19) of Ref. \cite%
{Okninski2015b}), in the interacting case, in form: 
\begin{equation}
\left. 
\begin{array}{l}
\pi _{\ \ \dot{B}}^{A}\ \zeta _{A\dot{D}}=m\chi _{\dot{B}\dot{D}}\smallskip
\\ 
\pi _{A}^{\ \ \dot{D}}\ \chi _{\dot{B}\dot{D}}=-m\zeta _{A\dot{B}}%
\end{array}
\right\}  \label{HH1}
\end{equation}
and it is assumed that 
\begin{equation}
\chi _{\dot{B}\dot{D}}=\chi _{\dot{D}\dot{B}}  \label{s=1}
\end{equation}
what is the $s=1$ constraint. In Eqs. (\ref{HH1}) we have $\pi ^{A\dot{B}
}=\left( \sigma ^{0}\pi ^{0}+\overrightarrow{\sigma }\cdot \overrightarrow{
\pi }\right) ^{A\dot{B}}$, $\pi ^{\mu }=p^{\mu }-qA^{\mu }$, $\sigma ^{k}$ $%
\left( k=1,2,3\right) $ are the Pauli matrices, and $\sigma ^{0}$ is the $%
2\times 2$ unit matrix. Let us note that equations (\ref{HH1}), (\ref{s=1}),
which can be written in the $7\times 7$\ Hagen-Hurley form, were first
proposed by Dirac \cite{Dirac1936}.

Equations (\ref{HH1}) in explicit form read: 
\begin{subequations}
\label{HH2}
\begin{eqnarray}
&&\left. 
\begin{array}{rcr}
-\left( \pi ^{1}+i\pi ^{2}\right) \chi _{\dot{1}\dot{1}}-\left( \pi ^{0}-\pi
^{3}\right) \chi _{\dot{2}\dot{1}} & = & -m\zeta _{1\dot{1}} \\ 
\left( \pi ^{0}+\pi ^{3}\right) \chi _{\dot{1}\dot{1}}+\left( \pi ^{1}-i\pi
^{2}\right) \chi _{\dot{2}\dot{1}} & = & -m\zeta _{2\dot{1}} \\ 
-\left( \pi ^{1}-i\pi ^{2}\right) \zeta _{1\dot{1}}-\left( \pi ^{0}-\pi
^{3}\right) \zeta _{2\dot{1}} & = & m\chi _{\dot{1}\dot{1}} \\ 
\left( \pi ^{0}+\pi ^{3}\right) \zeta _{1\dot{1}}+\left( \pi ^{1}+i\pi
^{2}\right) \zeta _{2\dot{1}} & = & m\chi _{\dot{2}\dot{1}}%
\end{array}
\right\}  \label{HH2a} \\
&&\left. 
\begin{array}{rcr}
-\left( \pi ^{1}+i\pi ^{2}\right) \chi _{\dot{1}\dot{2}}-\left( \pi ^{0}-\pi
^{3}\right) \chi _{\dot{2}\dot{2}} & = & -m\zeta _{1\dot{2}} \\ 
\left( \pi ^{0}+\pi ^{3}\right) \chi _{\dot{1}\dot{2}}+\left( \pi ^{1}-i\pi
^{2}\right) \chi _{\dot{2}\dot{2}} & = & -m\zeta _{2\dot{2}} \\ 
-\left( \pi ^{1}-i\pi ^{2}\right) \zeta _{1\dot{2}}-\left( \pi ^{0}-\pi
^{3}\right) \zeta _{2\dot{2}} & = & m\chi _{\dot{1}\dot{2}} \\ 
\left( \pi ^{0}+\pi ^{3}\right) \zeta _{1\dot{2}}+\left( \pi ^{1}+i\pi
^{2}\right) \zeta _{2\dot{2}} & = & m\chi _{\dot{2}\dot{2}}%
\end{array}
\right\}  \label{HH2b}
\end{eqnarray}
where the condition $\chi _{\dot{B}\dot{D}}=\chi _{\dot{D}\dot{B}}$ is not
imposed. We thus get two Dirac equations or, alternatively, a single Dirac
equation with generalized solution $\mathbb{B}=\left( 
\begin{array}{cc}
\zeta _{1\dot{1}} & \zeta _{1\dot{2}} \\ 
\zeta _{2\dot{1}} & \zeta _{2\dot{2}} \\ 
\chi _{\dot{1}\dot{1}} & \chi _{\dot{1}\dot{2}} \\ 
\chi _{\dot{2}\dot{1}} & \chi _{\dot{2}\dot{2}}%
\end{array}
\right) $ 
\end{subequations}
\begin{equation}
\left( \pi ^{0}\gamma ^{0}-\pi ^{1}\gamma ^{1}-\pi ^{2}\gamma ^{2}-\pi
^{3}\gamma ^{3}\right) \mathbb{B}=m\mathbb{B},  \label{Dirac-s=1,b}
\end{equation}
generalizing Eq. (24) of Ref. \cite{Okninski2015b}.

\section{Decay of spin $1$ bosons}

\label{betaW}

We note that solutions of two Dirac equations (\ref{HH2}) are non-standard
since they involve higher-order spinors rather than spinors $\xi _{A}$,$\
\eta _{\dot{B}}$. To interpret Eqs. (\ref{HH2}) we put: 
\begin{subequations}
\label{SUB}
\begin{eqnarray}
\chi _{\dot{B}\dot{D}}\left( x\right) &=&\eta _{\dot{B}}\left( x\right)
\alpha _{\dot{D}}\left( x\right)  \label{sub1} \\
\zeta _{A\dot{B}}\left( x\right) &=&\xi _{A}\left( x\right) \alpha _{\dot{B}%
}\left( x\right)  \label{sub2}
\end{eqnarray}%
where $\alpha _{\dot{A}}\left( x\right) $ is the Weyl spinor while $\eta _{%
\dot{B}}\left( x\right) $, $\xi _{A}\left( x\right) $ are the Dirac spinors.
Note that now $\chi _{\dot{1}\dot{2}}\neq \chi _{\dot{2}\dot{1}}$ and,
accordingly, the spin is not determined -- more exactly, the spin is in the $%
0\oplus 1$ space. It means that we consider virtual (off-shell) bosons. This
substitution is in the spirit of the method of fusion of de Broglie \cite%
{deBroglie1943,Corson1953} (similar ansatz was used in the $s=0$ case \cite%
{Okninski2015a}). After the substitution of (\ref{SUB}) into Eqs. (\ref{HH2}
) we obtain two equations:

\end{subequations}
\begin{equation}
\left. 
\begin{array}{rcl}
-\left( \pi ^{1}+i\pi ^{2}\right) \eta _{\dot{1}}\alpha _{\dot{A}}-\left(
\pi ^{0}-\pi ^{3}\right) \eta _{\dot{2}}\alpha _{\dot{A}} & = & -m\xi
_{1}\alpha _{\dot{A}} \\ 
\left( \pi ^{0}+\pi ^{3}\right) \eta _{\dot{1}}\alpha _{\dot{A}}+\left( \pi
^{1}-i\pi ^{2}\right) \eta _{\dot{2}}\alpha _{\dot{A}} & = & -m\xi
_{2}\alpha _{\dot{A}} \\ 
-\left( \pi ^{1}-i\pi ^{2}\right) \xi _{1}\alpha _{\dot{A}}-\left( \pi
^{0}-\pi ^{3}\right) \xi _{2}\alpha _{\dot{A}} & = & m\eta _{\dot{1}}\alpha
_{\dot{A}} \\ 
\left( \pi ^{0}+\pi ^{3}\right) \xi _{1}\alpha _{\dot{A}}+\left( \pi
^{1}+i\pi ^{2}\right) \xi _{2}\alpha _{\dot{A}} & = & m\eta _{\dot{2}}\alpha
_{\dot{A}}%
\end{array}
\right\}  \label{Dirac1}
\end{equation}
where $\dot{A}=\dot{1},\dot{2}$, and, after substituting solution of the
Weyl equation 
\begin{equation}
p^{A\dot{B}}\alpha _{\dot{B}}=0,  \label{Weyl}
\end{equation}
$\alpha _{\dot{A}}\left( x\right) =\hat{\alpha}_{\dot{A}}e^{ik\cdot x}$, $%
k^{\mu }k_{\mu }=0$, we get a single Dirac -- equation for spinors $\xi
_{A}\left( x\right) $, $\eta _{\dot{B}}\left( x\right) $: 
\begin{equation}
\left. 
\begin{array}{rcl}
-\left( \tilde{\pi}^{1}+i\tilde{\pi}^{2}\right) \eta _{\dot{1}}-\left( 
\tilde{\pi}^{0}-\tilde{\pi}^{3}\right) \eta _{\dot{2}} & = & -m\xi _{1} \\ 
\left( \tilde{\pi}^{0}+\tilde{\pi}^{3}\right) \eta _{\dot{1}}+\left( \tilde{
\pi}^{1}-i\tilde{\pi}^{2}\right) \eta _{\dot{2}} & = & -m\xi _{2} \\ 
-\left( \tilde{\pi}^{1}-i\tilde{\pi}^{2}\right) \xi _{1}-\left( \tilde{\pi}
^{0}-\tilde{\pi}^{3}\right) \xi _{2} & = & m\eta _{\dot{1}} \\ 
\left( \tilde{\pi}^{0}+\tilde{\pi}^{3}\right) \xi _{1}+\left( \tilde{\pi}
^{1}+i\tilde{\pi}^{2}\right) \xi _{2} & = & m\eta _{\dot{2}}%
\end{array}
\right\}  \label{Dirac2}
\end{equation}
where $\tilde{\pi}^{\mu }\equiv \pi ^{\mu }+k^{\mu }$, since components $%
\alpha _{\dot{1}}\left( x\right) $, $\alpha _{\dot{2}}\left( x\right) $
cancel out.

Equations (\ref{Weyl}), (\ref{Dirac2}) describe two spin $\frac{1}{2}$
particles, whose spins can couple to $s=0$ or $s=1$, i.e. $\frac{1}{2}
\otimes \frac{1}{2}=0\oplus 1$.

\section{Conclusions}

\label{conclusions}

Results obtained in Sections \ref{generalized}, \ref{betaW} cast new light
on the Hagen-Hurley equations as well as on weak decays of spin $1$ bosons.
We have shown that transition from equation (\ref{HH1}), describing a spin $%
s=1$ particle, to equations (\ref{Weyl}), (\ref{Dirac2}), via substitution (%
\ref{SUB}) -- which means that now $s\in 0\oplus 1$, corresponds to decay of
this particle into a Weyl antineutrino, cf. Eq. (\ref{Weyl}), and a Dirac
lepton, cf. Eq. (\ref{Dirac2}). Indeed, it should be a weak decay since Eq. (%
\ref{HH1}) violates parity. The spin of this particle becomes undetermined in the process of decay, 
more exactly it belongs to the $0\oplus 1$ space -- this suggests that this
is a virtual particle. Therefore, the products, a lepton and a antineutrino,
should have total spin $0$ or $1$ and there should be a third particle to
secure spin conservation.

The above descritption fits a (three-body) beta decay with formation of a
virtual $W^{-}$ boson, decaying into a lepton and antineutrino. This is most
conveniently explained in the case of a mixed beta decay \cite{Krane1988}: 
\begin{equation}
n\left( \uparrow \right) \longrightarrow \left\{ 
\begin{array}{l}
p\left( \downarrow \right) +\left[ e\left( \uparrow \right) \bar{\nu}%
_{e}\left( \uparrow \right) \right] \qquad \text{Gamow-Teller
transition\smallskip } \\ 
p\left( \uparrow \right) +\left[ e\left( \uparrow \right) \bar{\nu}%
_{e}\left( \downarrow \right) \right] \qquad \text{Fermi transition}%
\end{array}%
\right.   \label{mixed-beta}
\end{equation}%
where products of the $W^{-}$ boson decay (see \cite{Olive2014}) are shown
in square brackets and $\left( \uparrow \right) $ denotes spin $\frac{1}{2}$
-- this seems to correspond well to the proposed transition from Eq. (\ref%
{HH1}) to Eqs. (\ref{Weyl}), (\ref{Dirac2}). Since spin of the products of
decay of the virtual $W^{-}$ boson belongs to the $0\oplus 1$ space, their
spin can be $s=0$ or $s=1$. Moreover, in the case of the Gamow-Teller
transition there must be a spin-flip in the decaying nucleon. Let us add
here, that in the reaction (\ref{mixed-beta}) some neutrons ($82$\%) decay
according to the Gamow-Teller mechanism while some ($18$\%) undergo the
Fermi transition \cite{Krane1988}. This mixed mechanism is explained by
decoupled spins of the just born products -- indeed, the condition $\chi _{%
\dot{1}\dot{2}}=\chi _{\dot{2}\dot{1}}$\ for the spinor $\chi _{\dot{A}\dot{B%
}}$, due to the substitution (\ref{sub1}), does not hold and spin of the
products is in the $0\oplus 1$ space.

It is now obvious that another set of $7\times 7$ equations, involving
spinor $\eta _{AB}$\ rather than $\chi _{\dot{A}\dot{B}}$, see Eq. (18) of
Ref. \cite{Okninski2015b}, describes a $\beta ^{+}$ decay with intermediate $%
W^{+}$ boson. Let us note finally, that kinematics of the neutrino appears
in the Dirac equation for the electron with arbitrary neutrino
four-momentum, suggesting a continuous distribution of neutrino energy.

\subsection*{Conflict of Interests}

The author declares that there is no conflict of interests regarding the
publication of this paper.

\end{document}